\documentclass[draftcls, onecolumn, 12pt]{IEEEtran}

\usepackage[cmex10]{amsmath}
\usepackage{cite}
\usepackage{amssymb}
\usepackage{color,array}
\usepackage{graphicx}
\usepackage{enumitem}
\usepackage{multirow}
\usepackage{caption}
\usepackage{graphicx}

\newtheorem{teorema}{Theorem}
\newtheorem{proposicao}{Proposition}
\newtheorem{remark}{Remark}

\newtheorem{definicao}{Definition}
\newtheorem{corolario}{Corollary}
\newtheorem{exemplo}{Example}

\newcommand{\F}{\mathbb{F}}

\begin{document}

    \title{Characterization of metrics induced by hierarchical posets}

\author{Roberto~Assis~Machado,~Jerry~Anderson~Pinheiro~and~Marcelo~Firer}


\maketitle

    \begin{abstract}
In this paper we consider metrics determined by hierarchical posets and give explicit formulae for the main parameters of a linear code: the minimum distance and the packing, covering and Chebyshev radii of a code. We also present  ten characterizations of hierarchical poset metrics, including new characterizations and simple new proofs to the known ones.
    \end{abstract}


\section{Introduction}

\IEEEPARstart{T}{he} study of metrics induced by posets, originally introduced in 1995, by Brualdi, Graves and Lawrence \cite{brualdi}, became an interesting and productive area of research, partially because a number of unusual properties arise in this context, including the intriguing relative abundance of MDS and perfect codes, noticed, for example, in  \cite{MDSposetcodes} and \cite{pperfeitocodes}. Moreover, the study of classical metric invariants of Coding Theory, when considering unusual distances, raises many questions regarding very traditional and known results.

Over the years, the following code-related properties were proven to hold when considering a metric determined by a hierarchical poset: (i) the weight enumerator of a code is completely determined by the weight enumerator of its dual code (MacWilliams-type Identity), \cite{classification:macwilliamsidentity}; (ii) a linear code determines an association scheme, \cite{classification:schemes}; (iii) isometric linear  isomorphism between codes may be extended to the entire space (MacWilliams Extension Theorem), \cite{extensiontheoremforfiniterings3}; (iv) the packing radius of a code is a function of its minimum distance, \cite{felix_decomposicao_canonica}. {These properties appear dispersed throughout the literature and were proved by using many different combinatorial and algebraic tools: characters, association schemes, matroids, etc.} 

In this work, we prove that the previous properties (among others) are actually characterizations of hierarchical posets, in the sense that they hold (for any linear code, if the case) if, and only if, the metric is determined by a hierarchical poset. For all those properties (including the known ones) we give simple and short proofs. The proofs are based on the existence of a canonical decomposition introduced by Felix and Firer \cite{felix_decomposicao_canonica} and on a simple counterexample. 
 
Since this work gives a unifying treatment to many different problems concerning  codes with poset metrics, we make some efforts to present the context in which some relevant results were originally developed and also  to explain (without proving) some key results. For this reason, we start with a relatively lengthy section (Section \ref{pre}) presenting the preliminaries concepts and introducing some basic examples that follow us all  along the work. In Section \ref{bas} we present and explain the main result concerning poset codes that will be used in the sequence, namely,  the existence of a  canonical form of a code. Finally, in Section \ref{car} we present the original results of this work: explicit formulae for the invariants of a code when considering a hierarchical poset metric and ten different characterizations of hierarchical posets given in terms of coding theory properties.


\section{Preliminaries}\label{pre} 

\subsection{Poset metrics}
Although the concept of order is much wider, in this text we consider only orders over finite sets. Inasmuch, without loss of generality, we let $[n]=\{1,2,\ldots ,n\}$ be a finite set. We say that the pair $P=([n],\preceq_P)$ is a \emph{partially ordered set} (abbreviated as \emph{poset})  if $\preceq_P$ is a partial order relation on $[n]$, that is, for all $a,b,c\in [n]$ we have that: \emph{(i)} $a\preceq_Pa$; \emph{(ii)} $a\preceq_Pb$ and $b\preceq_Pc$ implies $a\preceq_Pc$; \emph{(iii)} $ a\preceq_Pb$ and $b\preceq_Pa$ implies $a=b$.  

An \textit{ideal} in a poset $P=([n],\preceq_P)$ is a 
 subset  $I\subseteq [n]$ such that, given $a\in [n]$ and $b\in I$, if $a \preceq_P b$, then $a\in I$.  Given 
$A\subseteq [n]$, we denote by $\langle A\rangle_P$ the smallest ideal of $P$ containing $A$ and call it the \emph{ ideal generated by} $A$. An ideal $\langle\{a\}\rangle_P$  generated by a set $A=\{a\}$ with a single element (a \emph{singleton}) is called a \emph{prime ideal}. For simplicity we denote $\langle a\rangle_P=\langle\{a\}\rangle_P$.
An element $a$ of an ideal $I\subseteq [n]$ is called a \emph{maximal} element of $I$ if $a\preceq_P x$ for some $x\in I$ implies $x=a$. The set of all maximal elements of an ideal $I$ is denoted by $\mathcal{M}_P(I)$. It is easy to see that, given an ideal $I\subseteq [n]$,  $\mathcal{M}_P(I)$ is the minimal set such that $\langle \mathcal{M}_P(I)\rangle_P=I$.   We remark that an ideal is prime if, and only if, it contains only one maximal element. Furthermore, this maximal element is also the generator of the ideal. 

We say that $b$ \emph{covers} $a$ if $a\preceq b$, $a\neq b$ and there is no extra element $c\in [n]$ such that $a\preceq_Pc\preceq_P b$. When considering ``small'' posets, i.e., small values of $n$, a good way to figure out the properties of the poset is a visual representation, called the \emph{Hasse diagram} of a poset $P = ([n], \preceq_P)$. The Hasse diagram is a directed graph whose vertices are the elements of $[n]$ and an edge connects $b$ to $a$ if and only if $b$ covers $a$. When drawing it on a paper, we assume that $b$ is ``above'' $a$ if $b$ covers $a$ so that the direction is always assumed to be downwards. In Table \ref{tab1} we picture the Hasse diagram of some posets over $[3]$.

\begin{table}[h]	
\centering
	\noindent\begin{tabular}{|c|c|} 
	\hline
	\textbf{Order Relation} & \textbf{Hasse Diagram} \\
	\hline 
	 \hline
	 & \multirow{3}{*}{\includegraphics{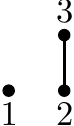}}\\
	 $2\preceq_{P_0} 3$ & \\
	  & \\
	 \hline
	  & \multirow{3}{*}{\includegraphics{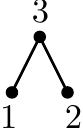}}\\
	  $1\preceq_{P_1} 3$ and $2\preceq_{P_1} 3$ & \\
	   & \\
	 \hline
	  & \multirow{3}{*}{\includegraphics{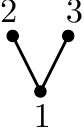}} \\
	  $1\preceq_{P_2} 2$ and $1\preceq_{P_2} 3$  & \\
	   & \\
	 \hline
	\end{tabular}	 
	 \noindent\begin{tabular}{|c|c|}
	 \hline
	\textbf{Order Relation} & \textbf{Hasse Diagram} \\
	\hline \hline
	 & \multirow{3}{*}{\includegraphics{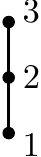}}\\
	  $1\preceq_{P_3} 2$, $1\preceq_{P_3} 3$ and $2\preceq_{P_3} 3$ & \\
	   & \\
	\hline
	 No relation (anti-chain) & \includegraphics{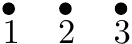} \\

	 \hline
	\end{tabular}
	 \caption{The Hasse diagrams of posets over $[3]$.}
	 \label{tab1}
\end{table}


The posets pictured in the table  are all hierarchical, except for the poset $P_0$, which will play an important role in this work, since it is a counterexample to many properties dealt with here, so we will call it \textit{Tiny Counterexample}.  A rigorous definition of poset isomorphism is postponed to Section \ref{bas}, but it is worth to note that, up to isomorphism, these are all the posets over $[3]$. 

\bigskip
In the context of coding theory, a poset over $[n]$ enables us to define a metric which may play a role similar to that of the usual Hamming metric. Let $\F_q^n$ be  an $n$-dimensional vector space over the finite field $\F_q$. Given $u\in \F_q^n$, the \textit{support} and the $P$\emph{-weight} of $u$ are defined respectively as
\[\mathrm{supp}(u)=\{i\in[n]:u_i\neq 0\}\]
and
\[\mathrm{wt}_P(u)=|\langle \mathrm{supp}(u)\rangle_P|,\]
where $|\cdot|$ denotes the cardinality of the given set. For simplicity, the set of maximal elements in the ideal generated by $\mathrm{supp}(u)$ is denoted as $\mathcal{M}_P(u)$. For $u,v\in \F_q^n$,
\[d_{P}(u,v)=\mathrm{wt}_P(u-v)\]
defines a metric over $\F_q^n$ called the \textit{poset metric}, or just the $P$\textit{-distance} between $u$ and $v$. {The space $\mathbb{F}_q^n$, when endowed with a poset metric $d_P$, is called a $P$\emph{-space}. Considering the five posets presented at Table \ref{tab1} and the vector $u=011\in\mathbb{F}_q^3$, we have that $\mathrm{wt}_{P_1}(u)=\mathrm{wt}_{P_2}(u)=\mathrm{wt}_{P_3}(u)=3$ and $\mathrm{wt}_P(u)=2$ in the other cases.
	 

An \textit{anti-chain} is a poset $P$ with a minimal set of relations, i.e., for any $a\neq b \in [n]$ neither $a \preceq_P b$ nor $b \preceq_P a$. Considering an anti-chain poset, we have that $\langle \mathrm{supp}(u)\rangle_P=\mathrm{supp}(u)$, hence it induces the well-known Hamming metric.
Besides the anti-chain poset, there is another poset that can also be considered as an extremal one, the one having a maximal number of relations, the \textit{chain poset}. In this case,
	 any two elements of $[n]$ are \emph{comparable} (or \emph{related}), i.e., given $a,b\in [n]$, either $a\preceq_Pb$ or $b\preceq_Pa$. The poset $P_3$ of Table \ref{tab1} is an example of a chain over $[3]$. Those two posets, the anti-chain and the chain, being posets determined by a minimal or maximal number of relations, also gives rise to metrics that, in some sense, are extremal ones. The Hamming metric, determined by an anti-chain, is the discrete counterpart of the Euclidean space, which models our sensorial perception of the world. The chain poset, on the other hand, gives rise to a very different type of metric, known as an ultra-metric. In an ultra-metric, the triangular inequality ($ d(x,z)\leq d(x,y)+d(y,z) $) is exchanged by the much stronger ultra-metric inequality ($ d(x,z)\leq \max {d(x,y),d(y,z)} $). This condition has a strong impact on the metric that appears in many places. Just as an example, considering a metric determined by a chain, the formula for the packing radius of a code is $ d(\mathcal{C})-1$ and not the usual $ \lfloor \frac{d(\mathcal{C})-1}{2}\rfloor$, where $d(\mathcal{C})$ is the minimum distance (see \cite{pperfeitocodes} for details). 
	 
	  It is worth noting that despite the fact that the metric determined by a chain defies our intuition, it is actually a very simple setting and the geometry of codes under this metric was described in details in \cite{panek2010classification}, including the characterization of perfect and MDS codes. 
	
	In the general case, the behaviors determined by chains and anti-chains are mixed together,  and computing the geometric invariants of a code becomes a very hard task, so the researchers started to consider different ways to combine chains and anti-chains. By performing disjoint unions of finite chains or by (hierarchically) relating families of anti-chain posets, we obtain the two most studied family of poset metrics: the Niederreiter-Rosembloom-Tsfasman (NRT) and hierarchical families of posets. 
	
	The NRT metrics are determined by a poset which is a disjoint union of chains. They have been widely investigated in the literature, see \cite{park2011ordered}, \cite{park2010linear}, \cite{ozen2004structure} and \cite{sharma2014macwilliams}. Its role as a metric model of a channel is well established, while the geometry of the space and the metric parameters of a code are not yet understood.
	
	The hierarchical posets, which are the subject of study of this work, are another possible generalization of both chain and anti-chain. Before we define what a hierarchical poset is, as an ``appetizer'',  we can say that a metric determined by a hierarchical poset is the true generalization of the Hamming metric in many different aspects. Just as an example, the only posets metrics where the minimum distance of a code determines its packing radius are the ones induced by hierarchical posets. Let us now introduce some concepts needed to define hierarchical posets.

\medskip
	 
	 The \emph{height}  $h(a)$ of an element $a \in P$ is the cardinality of a largest chain having  $a$ as the maximal element. The \emph{height} $h(P)$ of the poset is the maximal height of its elements, i.e., $h(P)=\max\left\{h(a):a\in [n]\right\}  $. The $i$-th level $\Gamma_i^P$ of a poset $P$ is the set of all elements with height $i$, i.e.,
	 \[
	 \Gamma_i^P = \{a\in [n] : h(a)=i\}.
	 \]
	 We stress that each level of a poset has the order structure of an anti-chain. On Table \ref{tab1}, the posets $P_0,P_1$ and $P_2$ (on the left side) all have $2$ levels, while on the right side we have a chain (with $3$ levels) and an anti-chain with a single level.
	 \begin{definicao}
	 A poset $P=([n], \preceq_P)$ is said to be \emph{hierarchical} \index{hierarchical poset} if elements at different levels (anti-chains) are always comparable, i.e., if $a\in \Gamma_i^P$ and $b\in \Gamma_{j}^P$, then $a\preceq_P b$ if, and only if, $ i <j $.	A \emph{hierarchical space} is a $P$-space, with $P$ a hierarchical poset.
	 \end{definicao}

Consider the posets $P$ and $Q$ over $[4]$ determined by the  Hasse diagrams on Figure \ref{eq007}: $P$ is hierarchical and $Q$ is not, because $1\in \Gamma_{1}^Q$, $4\in\Gamma_{2}^Q$ but $1$ and $4$ are not related (in $Q$).



\begin{figure}[h]
	\centering
	\begin{tabular}{cccc} 
	 & \multirow{3}{*}{\includegraphics{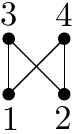}} & &\multirow{3}{*}{\includegraphics{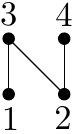}}\\
	 $P$ \ \ & & \ \ \ \ \ \ \ \  \ \ \ \ \ \ \ \   $Q$ \ \  &
	\end{tabular}
\caption{$P$ is hierarchical, $Q$ is not hierarchical}
	\label{eq007}
\end{figure}


From here on, if no confusion may arise, we will omit $P$ as a superscript in $\Gamma_i^P$ and as a subscript in $\preceq_P$, $\langle\cdot \rangle_P$, $\mathrm{wt}_P$ and $\mathcal{M}_P$.

\section{Basic properties of poset spaces and poset codes}\label{bas}

One of the most basic and initial results in coding theory is the  fact that, given a linear code $\mathcal{C} \subseteq\mathbb{F}_q^n$ there is an \emph{equivalent} code $\mathcal{C} ^{\prime}$ that has a generator matrix in a \emph{standard} form. To achieve the standard form we need to use basic operations on rows   to obtain a row-echelon form of the generator matrix, followed by a permutation of the columns. We should remark that basic operations on rows preserve the code, just giving another basis for it; while permutation of columns, being symmetries of the space relative to the Hamming metric, gives an equivalent (not necessarily equal) code. 

When considering metrics determined by hierarchical posets, it was shown in \cite{felix_decomposicao_canonica} that generator matrices can be characterized by a form similar to the standard one, 
known as \emph{canonical-systematic form}. The important part of this characterization is the canonical part, so we refer to it as the \textit{canonical form} of a code. In order to describe it, we will first characterize the group of linear isometries of a $P$-space, which was completely described in \cite{panek_group_of_linear_isometries}. Let us start with some definitions.

A map $T:\mathbb{F}_q^n \rightarrow \mathbb{F}_q^n$ is called a $P$\emph{-isometry} if $d_P(T(u),T(v)) =d_P(u,v)$ for every $u,v\in \F_q^n$. We denote by $GL_P(\mathbb{F}_q)$ the group of linear isometries of a $P$-space, i.e., 
\[GL_P(\mathbb{F}_q):= \{T:\mathbb{F}_q^n \rightarrow \mathbb{F}_q^n : T \mbox{\upshape{ is a linear }} P\mbox{\upshape{-isometry}}\}.\]
Two linear codes $\mathcal{C},\mathcal{C}^{\prime}\subseteq \F_q^n$ are said to be \emph{$P$-equivalent} if there is $T\in GL_P(\mathbb{F}_q)$ such that $T(\mathcal{C})=\mathcal{C}^{\prime}$. This definition agrees with the usual equivalence of codes in the Hamming setting.  We stress that two $P$-equivalent codes are, geometrically speaking, indistinguishable.  

Let $P$ and $Q$ be two posets over $[n]$. Given $X,Y\subseteq [n]$, an \textit{order isomorphism} (according to $P$ and $Q$) between $X$ and $Y$ is a bijection $\phi :X\rightarrow Y$ satisfying $i\preceq_P j$ if, and only if, $\phi(i)\preceq_Q\phi(j)$ for every $i,j\in X$. When $X=Y$ and $P=Q$, we call $\phi$ a $P$\textit{-automorphism}. We denote by $Aut(P)$ the group of automorphisms of the poset $P$.  Note that two isomorphic posets can always be represented by the same Hasse diagram, hence the posets $P$ and $Q$ of Figure \ref{eq007} are not isomorphic.


For $j\in [n]$, let $e_j\in\mathbb{F}_q^n$ be the vector whose $j$-th coordinate is equal to $1$ and $\mathrm{supp}(e_j)=j$. We denote by $\beta_n=\{e_1,\ldots,e_n\}$  the usual basis of $\mathbb{F}_q^n$. Given $T\in GL_P(\mathbb{F}_q)$, in \cite{panek_group_of_linear_isometries} it was proved that the ideal generated by $\textrm{supp}(T(e_j))$ is prime. This implies that $T\in GL_P(\mathbb{F}_q)$ determines a map $\Phi_T:[n]\rightarrow [n]$ mapping  $j\in[n]$ into $\mathcal{M}(T(e_j))$ which, a posteriori, happens to be a  $P$-automorphism.
Actually, it also works on the other way: given $\phi\in Aut(P)$, the linear map $T_{\phi}$ defined by $T_{\phi}(e_i)=e_{\phi (i)}$ is an element of $GL_P(\mathbb{F}_q)$. This leads to a characterization of the linear isometries of a poset space: every isometry $T\in GL_P(\mathbb{F}_q)$ is the product $T=T_{\phi}\circ T_U$ where $T_{\phi}$ is induced by an automorphism $\phi$ and $T_U$ is the linear map determined by 
\[T_U(e_j) = \sum_{i\preceq _Pj} u_{ij}e_i \text{ 
	with } u_{jj}\neq 0  , \text{ for all } j\in[n].\] 
We use the notation $T_U$ to indicate that the transformation is determined by an $n\times n$ matrix $U=(u_{ij})$, with  $u_{jj}\neq 0$ for all $ j\in[n] $ and $u_{ij}=0$ if $i\notin \langle j\rangle_P$.  We summarize the main results of \cite{panek_group_of_linear_isometries} (described in the previous paragraphs) in the next theorem.

\begin{teorema}\label{automorfismo_induzidopor_isometria}\label{idealprimo}
	Let $P=([n],\preceq)$ be a poset. Then the following three conditions hold:
	
	\begin{itemize}
		\item[(i)] The ideal $\langle \mathrm{supp}(T (e_i ))\rangle$ is prime for every $T\in GL_P(\mathbb{F}_q)$ and every $i \in [n]$;
		\item[(ii)] If $T\in GL_P(\mathbb{F}_q)$, then the map $\Phi_T: P \rightarrow P$ given by 
		\[\Phi_T(i) = \mathcal{M}\left( T(e_i)\right)\]
		is a $P$-automorphism.
		\item[(iii)] $T\in GL_P(\mathbb{F}_q)$ if, and only if, 
		\[T(e_j) = \sum_{i\preceq_P j} u_{ij}e_{\Phi_T(i)},\]
		where $\Phi_T$ is a $P$-automorphism (a posteriori defined as in item (ii)) and $u_{ij}$ are scalars with $u_{jj}\neq 0$ for every $j\in [n]$.
	\end{itemize}
\end{teorema} 

The canonical form is a key result to this work, so we shall outline  the main steps used in its construction.

Let $P$ be a poset over $[n]$ and let $\mathcal{C}\subseteq\mathbb{F}_q^n$ be a $k$-dimensional linear code with generator matrix $G$. We assume that the $i$-th level $\Gamma_i$ has $n_i$ elements and we write $N_1=0$ and  $N_i=\sum_{j=1}^{i-1}n_j$ for $1<i\leq h(P)$. Without loss of generality (up to an isomorphism of posets), we may assume that  the $n_i$ columns of $G$ labeled by $N_i+1 \leq j \leq N_{i+1}$ correspond to the elements of the $i$-th level $\Gamma_i$ of $P$. Furthermore, since elementary operations with rows does not change the code, we may assume that $G$ is in a ``lower triangular'' form and each \textit{pivot} $g_{ir_i}$ is the unique non-null entry in the $r_i$-th column of the matrix:
\[
G=\left( \begin{matrix}
 g_{11}&\cdots   &g_{1r_1}  &0            &  \cdots     &  0       & 0      & \cdots&        0 & 0       \cdots  0      \\ 
g_{21} & \cdots  &0  &g_{2(r_1+1)} &\cdots  & g_{2r_2} & 0           & \cdots& 0        & 0      \cdots   0     \\ 
\vdots & \vdots        &  0     &  \vdots     &\vdots  & 0  &\vdots       & \cdots&  0  &         \\ 

g_{k1} & \cdots  & 0 & g_{k(r_1+1)}& \cdots & 0 & g_{k(r_2+1)}& \cdots& g_{kr_k} & 0      \cdots  0
\end{matrix}  \right) .
\]
In addition, given a vector $u\in\mathbb{F}_q^n$, it follows from item (iii) of Theorem \ref{automorfismo_induzidopor_isometria}, that it  is possible to construct an isometry $T_U$  such that the support of $T_U(u)$ contains only the maximal elements of the support of $u$, i.e., $\mathrm{supp}(T_U(u))=\mathcal{M}(u)$. We call the vector $(T_U(u))$ the \emph{cleaned form} of $u$. 

 
Now, we use the pivot $g_{ir_i}$ of the $i$-th row to clean the vector corresponding to this row. Since we are using pivots to obtain the cleaned form of the rows, the cleaning operation in a row does not ``spoil'' the cleaned ones. By doing so, we get a matrix $\widetilde{G}$ that is a generator matrix of a code $\widetilde{\mathcal{C}}$ which is $P$-equivalent to $\mathcal{C}$ (since $\widetilde{G}$ was obtained from $G$ by considering a sequence of linear isometries). The matrix $\widetilde{G}=(\widetilde{g}_{ij})_{i=1,\ldots ,k,j=i,\ldots ,n}$ has the property that $j\preceq_Pr_i$ or $j>r_i$ implies $\widetilde{g}_{ij}=0$. 
	
Just now we shall assume that $P$ is a hierarchical poset. By doing so, we get that $\widetilde{g}_{ij}=0$ if $j \leq N_l <r_i $ for some $l$. This happens because  $j\leq N_l<r_i$ means that $j$ belongs to a level at most equal to $l$ and $r_i$ belongs to a level at least equal to $l+1$ and, assuming $P$ to be hierarchical, this means that $j\preceq_P r_i$.


The matrix $\tilde{G}$, formed by cleaned vectors, has a block diagonal structure
\[
\widetilde{G}=\left( \begin{matrix}
0 \cdots  0& \widetilde{G}_{1r_1}& 0 \cdots 0 &        0 & 0\cdots0  &\cdots&0   & 0 \cdots 0 \\ 
0 \cdots 0 & 0       & 0\cdots 0  &\widetilde{G}_{2r_2}  & 0\cdots 0 &\cdots&0   &  0 \cdots 0\\ 
\vdots     & \vdots  & \vdots     &  \vdots        &  \vdots   &\vdots&       0 & \vdots\\ 
0 \cdots 0 & 0       & 0\cdots0   & 0   & 0\cdots0  &\cdots&\widetilde{G}_{tr_t} &0 \cdots 0
\end{matrix}  \right)
\]
where each $ \widetilde{G}_{ir_i} $ is a $k_i\times n_{r_i}$ matrix, corresponding to the columns on the level $\Gamma_{r_i}$ of $P$.  
	 This is called a \emph{canonical form} of the generator matrix, as introduced in \cite{felix_decomposicao_canonica}.

The special form of $\widetilde{G}$, ensured by the hierarchical structure of $P$, has a striking consequence, that lays in the kernel of this work.  It may be better grasped when restated in terms of codes, instead of generator matrices. Consider the submatrix of $\widetilde{G}$ consisting of the rows that contains $\widetilde{G}_{ir_i}$. This matrix generates a code $\widetilde{\mathcal{C}}_{r_i}$ such that  $\mathrm{supp}(\widetilde{\mathcal{C}}_{r_i})\subseteq \Gamma_{r_i}$ for every $i\in [t]$. So, the code $\widetilde{\mathcal{C}} $ generated by $\widetilde{G}$, which is $P$-equivalent to $\mathcal{C}$, can be expressed as 
\begin{equation}\label{canonical}
\widetilde{ \mathcal{C}}=\widetilde{ \mathcal{C}}_1\oplus \widetilde{ \mathcal{C}}_2 \oplus \cdots \oplus \widetilde{ \mathcal{C}}_{h(P)},
\end{equation}
 where $\mathrm{supp}(\widetilde{ \mathcal{C}}_i)\subseteq \Gamma_i$. We note that we may have $\widetilde{ \mathcal{C}}_i=\{0\} $, or equivalently, $\mathrm{supp}(\widetilde{ \mathcal{C}}_i)=\emptyset$. This decomposition is called the $P$-\emph{canonical decomposition of} $\mathcal{C}$. We say that a code \emph{admits a} $P$-canonical decomposition if it is $P$-equivalent to a code that is decomposed as in (\ref{canonical}). We stress that, as previously outlined, in the hierarchical case every code admits a $P$-canonical decomposition.   
 
 \medskip
 	 We remark that, for a Hamming metric, which is induced by an anti-chain poset, every code is in its canonical decomposition, since the poset has a unique level. For other posets, it is not the case. 
  Just as an example, let $n=3$, and $\mathcal{C}=\{000,101\}\subseteq \mathbb{F}_2^3$.   Considering $P$ as either $P_1, P_2$ or $P_3$, we have that $\mathcal{C}$ is $P$-equivalent to $\widetilde{\mathcal{C}}=\{000,001\}  $ and since $\mathrm{supp} (\widetilde{\mathcal{C}})$ is a singleton, it is contained in a unique level, namely $\Gamma_2^{P_1}$, $\Gamma_2^{P_2}$ or $\Gamma_3^{P_3}$ respectively. However, considering the tiny counterexample $P_0$, {the unique non-identity element of $GL_{P_0}(\mathbb{F}_2) $ is the map $T(x_1e_1+x_2e_2+x_3e_3)=x_1e_1+(x_2+x_3)e_2+x_3e_3$ and since $T(101)=111$, it follows that  $\mathcal{C}$ does not admit a $P_0$-canonical decomposition. The existence of a $P$-canonical decomposition of any given code is a property shared only by hierarchical posets. 

		\begin{teorema}\label{thm:canonical_decomposition}
			A poset $P = ([n], \preceq)$ with $l$ levels is hierarchical if, and only if, any linear code $\mathcal{C}\subseteq \mathbb{F}_q^n$ admits a $P$-canonical decomposition.
		\end{teorema}
				\begin{IEEEproof}
			If $P$ is hierarchical, then the existence of a $P$-canonical decomposition follows from the construction just described, for more details, see \cite[Corollary 1]{felix_decomposicao_canonica}. 
			Suppose $P$ is not hierarchical and let $i\in[l]$ be the lowest level of $P$ for which there  are $a\in \Gamma_i$ and $b\in \Gamma_{i+1}$ such that $a\not\preceq b$, i.e., $b$ is not greater than $a$ according to $P$. The $1$-dimensional linear code $\mathcal{C} = span\{ e_a+e_b\}$ does not admit a  $P$-canonical decomposition. 
			Indeed, Theorem \ref{idealprimo} ensures that for any linear isometry $T\in GL_P(\mathbb{F}_q)$, the ideals $\langle \mathrm{supp}(T(e_a))\rangle$ and $\langle \mathrm{supp}(T(e_b))\rangle$ are both prime, generated by $\Phi_T(a)$ and $\Phi_T(b)$ respectively, where $\Phi_T$ is the $P$-automorphism described in Theorem \ref{automorfismo_induzidopor_isometria}.
			 Since $\Phi_T$ is a $P$-automorphism, it follows that $\Phi_T(a)\in \Gamma_i$ and $\Phi_T(b)\in\Gamma_{i+1}$. Moreover, since $a\not\preceq b$, we have that $\Phi_T(a)\not\preceq \Phi_T(b)$. It follows that $\mathcal{M}(T(span\{ e_a + e_b\})) = \{\Phi_T(a), \Phi_T(b)\}$ is not contained in a single level. Since $dim(\mathcal{C}) = 1$ and $T\in GL_P(\mathbb{F}_q)$ is arbitrary, we find that $\mathcal{C}$ does not admit a $P$-canonical decomposition.
		\end{IEEEproof} 
			\medskip
		

		Let $P$ be a poset over $[n]$ with $l$ levels. For every $i\in [l]$, if $n_i = |\Gamma_i|$ and 
\[
	\widetilde{\mathbb{F}_q^{n_i}} = \{u\in \mathbb{F}_q^n \ : \ \mathrm{supp}(u)\subseteq \Gamma_i\},
\]
then the entire space can be decomposed according to the levels of $P$, i.e., 
\[\mathbb{F}_q^n = \bigoplus_{i\in [l]} \widetilde{\mathbb{F}_q^{n_{i}}}=\widetilde{\mathbb{F}_q^{n_1}} \oplus \widetilde{\mathbb{F}_q^{n_2}}\oplus \cdots \oplus \widetilde{\mathbb{F}_q^{n_l}}.\]
We call this the $P$-\emph{level decomposition} of $\mathbb{F}_q^n$. By definition, the parcels $\widetilde{\mathcal{C}}_i$ of a $P$-canonical decomposition are subspaces of $\widetilde{\mathbb{F}_q^{n_i}}$. Since $\widetilde{\mathbb{F}_q^{n_i}} \subseteq \mathbb{F}_q^n$ is naturally isomorphic to $\mathbb{F}_q^{n_i}$, even though $\mathbb{F}_q^{n_i}\not\subseteq \mathbb{F}_q^{n}$, we will use $\mathbb{F}_q^{n_i}$ and $\widetilde{\mathbb{F}_q^{n_i}}$ interchangeable and, from here on,  write $\widetilde{\mathcal{C}}_i\subseteq \mathbb{F}_q^{n_i}$.
		

\section{Characterizations and metric invariants of codes according to hierarchical poset metrics}\label{car}


\subsection{Metric Invariants}

Many important properties of codes are determined by metric invariants. For example, the error correction capability of a code is determined by its packing radius, which in turn, may (or not) be determined by its minimum distance. In this section we will characterize invariants that are central to describe some important properties of codes. 

	Since the distance $d_P$ assumes only values in $[n]\cup \{0\}$,  for $u\in \F_q^n$ and $r\in [n]$, let 
	\[\mathbb{B}(u,r) = \{v\in \mathbb{F}_q^n : d_P(u,v)\leq r\}\]
	and
	\[\mathbb{S}(u,r) = \{v\in  \mathbb{F}_q^n : d_P(u,v) = r\}\]
	be the ball and sphere with radius $r$ and center $u$, respectively. If $\mathcal{S}$ is a subset of $\mathbb{F}_q^n$, then: 
	\begin{itemize}[align=right]
		\item[(i)] The \emph{Minimum distance} of $\mathcal{S}$ is the smallest distance between two distinct elements of $\mathcal{S}$, i.e.,
		$$d_P(\mathcal{S})=\min\{d_P(x,y):x,y\in \mathcal{S},x\neq y \};$$
		\item[(ii)] The \emph{Packing Radius} of  $\mathcal{S}$ is the largest positive integer $\mathcal{P}_P(\mathcal{S})$ such that the balls of radius $\mathcal{P}_P(\mathcal{S})$ centered at the elements of $\mathcal{S}$  are pairwise disjoint, i.e.,
		\[
		 	\mathcal{P}_P(\mathcal{S})=\max\{i\in\mathbb{Z} \ : \ \mathbb{B}(u,i)\cap\mathbb{B}(v,i)=\emptyset \ \forall \ u,v\in \mathcal{S} \text{ with } u\neq v\};
		 \] 
		\item[(iii)] The \emph{Covering Radius} of  $\mathcal{S}$ is the smallest positive integer $\mathrm{C}_{ov,P}(\mathcal{S})$ such that the balls of radius $\mathrm{C}_{ov,P}(\mathcal{S})$ centered at the elements of $\mathcal{S}$ cover $\F_q^n$, i.e.,
		\[
			\mathrm{C}_{ov,P}(\mathcal{S})=\min\left\{i\in\mathbb{Z} \ : \ \mathbb{F}_q^n= \bigcup_{u\in \mathcal{S}}\mathbb{B}(u,i) \right\};
		\]
		\item[(iv)] The \emph{Chebyshev Radius} of $\mathcal{S}$ is the smallest positive integer $\mathcal{R}_P(\mathcal{S})$ such that there is a ball centered in a vector $u\in\mathbb{F}_q^n$ with radius $\mathcal{R}_P(\mathcal{S})$ containing $\mathcal{S}$, i.e.,
		\[
			\mathcal{R}_P(\mathcal{S})=\min\{i\in\mathbb{Z} \ : \ \mathcal{S} \subseteq \mathbb{B}(u,i) \text{ for some } u\in\mathbb{F}_q^n\}.
		\]
		A vector $u$ reaching this minimum is called a \index{Chebychev center}\emph{Chebyshev center of} $\mathcal{S}$.
\end{itemize} 

If no confusion may arise, we may omit the index $P$ and write just $d(\mathcal{S}), \mathcal{P}(\mathcal{S}), C_{ov}(\mathcal{S})$ and $\mathcal{R}(\mathcal{S})$ for the minimum distance, packing, covering and Chebyshev radii, respectively. By using the $P$-canonical decomposition, in \cite{felix_decomposicao_canonica}, the authors gave explicit formulae for the minimum distance and the packing radius. We will describe those results and provide explicit formulae for the remaining invariants, the covering and Chebyshev radii. We stress that all the formulae are expressed in terms of the invariants obtained according to the Hamming metric considered on each component of a $P$-canonical decomposition of a code.

Before we give such explicit formulae, note that by suitable relabeling a poset $P$, we may assume that an ideal $I$ of $P$ with $|I|=s$ is, by itself, a poset over $[s]$. We call it the \emph{subposet} structure induced on $I$ by $P$. The next proposition ensures that the packing radius of a code $\mathcal{C}$ (the  most relevant of the metric invariants previously introduced),  depends only on the subposet structure of the ideal generated by the support of $\mathcal{C}$. 



\begin{proposicao}\label{just}
	Let $\mathcal{C}$ be a linear code and let $I= \langle \mathrm{supp} (\mathcal{C}) \rangle  $ be the ideal generated by the support of $\mathcal{C}$ and let $\widehat{\mathcal{C}}$ be the code obtained by puncturing $\mathcal{C}$ on $[n]\setminus I$. Then, the packing radius of $\widehat{\mathcal{C}}$ according to $I$ coincides with the packing radius of $\mathcal{C}$ according to $P$. 
\end{proposicao} 

\begin{IEEEproof}
	If $x\in \mathbb{F}_q^n$, $c\in\mathcal{C}$ and $r>0$, then 
	\[
	x\in B_P(0,r)\cap B_P(c,r) \Rightarrow \widehat{x}\in B_{I}(\widehat{0},r)\cap B_{I}(\widehat{c},r)
	\]
	where $s=| I|$, $\widehat{x}, \widehat{0} \in \mathbb{F}_q^s$ and $\widehat{c} \in \widehat{\mathcal{C}}$ are the punctured vectors of $x, \ 0$ and $c$ respectively. Hence, $\mathcal{R}_P(\mathcal{C})\geq \mathcal{R}_I(\widehat{\mathcal{C}})$. 
	
	Let $c\in\widehat{\mathcal{C}}$ and $y\in\mathbb{F}_q^{s}$.  Since $\mathrm{supp} (\mathcal{C})\subseteq I $, there is a unique   $c^\prime\in \mathcal{C}$ such that  $\widehat{c^\prime}=c$. We denote  by $y^\prime$ the unique vector of $\mathbb{F}_q^n$ satisfying $\mathrm{supp}(y)\subseteq I$ and $\widehat{y^\prime}=y$.  Consider  $\widehat{0} \in\mathbb{F}_q^s$ and  $0 \in\mathbb{F}_q^n$ the null vectors in the corresponding spaces. It follows  that 
	\[
	y\in B_I(\widehat{0},r)\cap B_I(c,r) \Rightarrow y^\prime\in B_P(0,r)\cap B_P(c^\prime,r),
	\]
	hence $\mathcal{R}_I(\widehat{\mathcal{C}})\geq \mathcal{R}_P(\mathcal{C})$.
\end{IEEEproof}

Considering the subposet structure induced on an ideal $I$ of $P$, it may happen that $P$ is not hierarchical, while  $I$ is so. Indeed, the tiny counterexample $P_0$ is not hierarchical but for any $I\subseteq [3]$, with $|I|=1$ or $|I|=2$, we have that the subposet induced on $I$ by $P_0$ is hierarchical.

\begin{exemplo}
	Let $P$ and $Q$ be the posets with Hasse diagrams depicted in Figure \ref{figs01}. If $\mathcal{C}=\{000000,101100\}$ then $I=\langle\mathrm{supp}(\mathcal{C})\rangle_P=\{1,2,3,4\}$. Hence, $Q=I$ and $\widehat{\mathcal{C}}=\{0000,1011\}$. Furthermore, $\mathcal{R}_P(\mathcal{C})=3=\mathcal{R}_Q(\mathcal{\widehat{C}})=\mathcal{R}_I(\mathcal{\widehat{C}})$.
	\begin{figure}[h]
	\centering
	\includegraphics[scale=1.3]{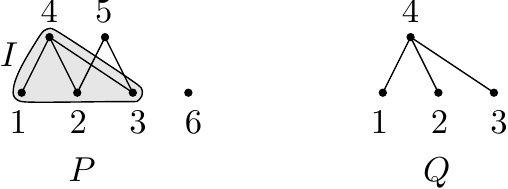}			
\caption{Hasse diagrams of P and Q.}
	\label{figs01}
\end{figure}
\end{exemplo}

Now we give explicit formulae for the main metric invariants of a code.

\begin{proposicao}\label{invariantesmetricos}Let $P = ([n], \preceq)$ be a hierarchical poset with $l$ levels and denote $n_i=|\Gamma_i|$. 
Let $\{0\}\neq\mathcal{C}\subseteq \mathbb{F}_q^n$ be a  linear code and $\mathcal{C}_1 \oplus \mathcal{C}_2 \oplus \cdots \oplus \mathcal{C}_l$  a $P$-canonical decomposition of $\mathcal{C}$. Then:
\begin{itemize}
\item[(i)] The minimum distance of $\mathcal{C}$ is given by
\[\displaystyle d_P(\mathcal{C}) =  \sum_{i=1}^{t_1-1} n_i + d_H(\mathcal{C}_{t_1})\mbox{\upshape{,}}\]
where $t_1 = \min\{i\in[l]: \mathcal{C}_i \neq \{0\}\}$ and $d_H(\mathcal{C}_{t_1})$ is the minimum distance of $\mathcal{C}_{t_1}$ considered as a code in the Hamming space $\mathbb{F}_q^{n_{t_1}}$;

\item[(ii)] The packing radius of $\mathcal{C}$ is given by
\[\displaystyle\mathcal{P}(\mathcal{C}) =  \sum_{i=1}^{t_1-1} n_i + \left\lfloor \frac{d_H(\mathcal{C}_{t_1})-1}{2}\right\rfloor\mbox{\upshape{;}}\]

\item[(iii)] The covering radius of $\mathcal{C}$ is given by
\[\displaystyle\mathrm{C}_{ov}(\mathcal{C}) =  \sum_{i=1}^{h-1} n_i + \mathrm{C}_{ov}^H(\mathcal{C}_{{h}})\mbox{\upshape{,}}\]
where: $\mathrm{C}_{ov}^H(\mathcal{C}_{{h}})$ is the covering radius of $\mathcal{C}_{{h}}$ considered as a code in the Hamming space $\mathbb{F}_q^{n_h}$;    $h = \min\{i\in[l-1]: \mathcal{C}_{j} = \mathbb{F}_q^{n_j}, \forall j >i\}$ if $\mathcal{C}_l = \mathbb{F}_q^{n_l}$ or $h=l$ if $\mathcal{C}_l \neq \mathbb{F}_q^{n_l}$;

\item[(iv)] The Chebyshev radius of $\mathcal{C}$ is given by
$$\displaystyle\mathcal{R}(\mathcal{C}) =  \sum_{i=1}^{r-1} n_i + \mathcal{R}^H(\mathcal{C}_{r})\mbox{\upshape{,}}$$
where $r = \max\{i\in[l]: \mathcal{C}_{i}\neq \{0\}\}$ and $\mathcal{R}^H(\mathcal{C}_{r})$ is the Chebyshev radius of $\mathcal{C}_{r}$ considered as a code in the Hamming space $\mathbb{F}_q^{n_r}$.
\end{itemize}
\end{proposicao}

\begin{IEEEproof}
	As a general remark for the proof, since all the radii and measures are invariant by any linear $P$-isometry, we may assume, without loss of generality, that $\mathcal{C}$ is a $P$-canonical decomposition of itself, i.e., $\mathcal{C}= \mathcal{C}_1 \oplus \mathcal{C}_2 \oplus \cdots \oplus \mathcal{C}_l$ and $\mathrm{supp}(\mathcal{C}_i)\subseteq \Gamma_i$.
	
For item (i), since $\mathcal{C} $ is linear, the minimum distance equals the minimum weight and since $P$ is hierarchical, the characterization of the minimum weight  follows immediately from the definition. Item (ii) was first proved in \cite{felix_decomposicao_canonica}, Propositions 3. The main idea is to proof that $\mathcal{P}(\mathcal{C})=\sum_{i=1}^{t_1-1}n_i +\mathcal{P}^H(\mathcal{C}_{t_1})$ where $\mathcal{P}^H(\mathcal{C}_{t_1})$ is the packing radius of $\mathcal{C}_{t_1}$ according to the Hamming metric on $\mathbb{F}_q^{n_{t_1}} $,  and, as is well known,  it is fully determined by $d_H(\mathcal{C}_{t_1})$: $\mathcal{P}^H(\mathcal{C}_{t_1})= \left\lfloor \frac{d_H(\mathcal{C}_{t_1})-1}{2}\right\rfloor $. We shall prove the last two items.

\begin{itemize}
	\item [(iii)] 
Suppose there is $u \in \mathbb{F}_q^n$ such that 
\[d_P(u,c) > \sum_{i=1}^{{h}-1} n_i + \mathrm{C}_{ov}^H(\mathcal{C}_{{h}})\]
for every $c\in\mathcal{C}$. Since $d_P(u,c) = \sum_{i=1}^{{h}-1} n_i + d_H(u_h, c_h)$ for every $c\in\mathcal{C}$ satisfying $\emptyset\neq\mathcal{M}(u-c)\subseteq \Gamma_h$, then $d_H(u_h, c_h) > \mathrm{C}_{ov}^H(\mathcal{C}_{{h}})$, which is a contradiction. Therefore 
\begin{equation*}
\displaystyle\mathrm{C}_{ov}(\mathcal{C}) \leq  \sum_{i=1}^{{h}-1} n_i + \mathrm{C}_{ov}^H(\mathcal{C}_{{h}})\mbox{\upshape{.}}
\end{equation*}

If the equality does not hold in the previous equation, then,  for each $u\in\mathbb{F}_q^n$, there is $c\in \mathcal{C}$ such that  $d_H(u_h, c_h)<\mathrm{C}_{ov}^H(\mathcal{C}_{{h}})$. This contradicts the minimality of $\mathrm{C}_{ov}^H(\mathcal{C}_{{h}})$.

\item[(iv)] Considering the Hamming metric over $\mathcal{C}_r$, let $u\in\mathbb{F}_q^{n_{r}} \subseteq\mathbb{F}_q^n$ be a Chebyshev center of $\mathcal{C}_r \subseteq \mathbb{F}_q^{n_{r}}$. Since $P$ is hierarchical, given $0\neq v\in\F_q^n$, we have that $\mathcal{M}(v)\subseteq \Gamma_i$ for some $i\in [l]$. The maximality of $r$ ensures that if $c\in \mathcal{C}$ then $\mathcal{M}(c)\subseteq \Gamma_j$ for some $j\leq r$ and since $\mathcal{M}(u)\subseteq \Gamma_r$, we get that $\mathcal{M}(u-c)\subseteq \Gamma_s$ for some $s\leq r$. Also, by the $P$-level decomposition of $\mathbb{F}_q^n$ and the canonical decomposition of $\mathcal{C}$, we may write $u=u_r$ and $c=c_1+\cdots+c_l$ where $c_i\in \mathcal{C}_i$. With this notation we have that 
\begin{align*}
 d_P(u,c) & =   n_1 + n_2 + \cdots + n_{s-1} + d_H(u_s, c_s) \\
&\leq n_1 + n_2 + \cdots + n_{r-1} + d_H(u_r, c_r)\\
& \leq n_1 + n_2 + \cdots + n_{r-1} + \mathcal{R}^H(\mathcal{C}_r),
\end{align*}
where the equality follows from the definition of the $P$-distance when $P$ is hierarchical. The first inequality follows from the fact that $s\leq r$  and the second inequality is a consequence of the definition of the Chebyshev radius. It follows that 
$$\mathcal{R}(\mathcal{C})\leq n_1 + n_2 + \cdots + n_{r-1} + \mathcal{R}^H(\mathcal{C}_r).$$ 

For the opposite inequality, let $v\in\mathbb{F}_q^n$ be the Chebyshev center of $\mathcal{C}$. Then $\mathcal{M}(v)\subseteq \Gamma_s$ for some $s\leq r$, since otherwise, by the maximality of $r$, $d_P(0,c)< d_P(v,c)=wt(v)$ for every $c\in \mathcal{C}$, i.e., $v$ would not be a Chebyshev center. The maximality of $r$ also ensures that for every $c\in\mathcal{C}$, there is  $s\leq r$ such that  $\mathcal{M}(v-c)\subseteq \Gamma_s$. 

Let $c_r \in \mathcal{C}_r$ be a codeword such that $d_H(v_r,c_r)$ is maximal. Since $\mathcal{C}_r \neq \{0\}$, there are at least two codewords in  $\mathcal{C}_r $ hence $d_H(v_r,c_r)>0$. 
 As a consequence, we have that $\mathcal{M}(v-c_r)\subseteq \Gamma_r$. The definition of the Chebyshev radius ensures that $d_P(v,c_r)\leq \mathcal{R}(\mathcal{C})$, and so
\[
d_P(v,c_r)  = n_1 + n_2 + \cdots + n_{r-1} + d_H(v_r, c_r)\leq\mathcal{R}(\mathcal{C})\mbox{\upshape{.}}
\]
Since $d_P(v_r,c_r)\geq d_P(v_r,c_r^\prime)$ for every $c_r^\prime \in \mathcal{C}_r$, we have that $\mathcal{R}^H(\mathcal{C}_r)\leq d_H(v_r,c_r)$. Therefore,  
\[
n_1 + n_2 + \cdots + n_{r-1} + \mathcal{R}^H(\mathcal{C}_r) \leq d_P(v,c_r) \leq\mathcal{R}(\mathcal{C})\mbox{\upshape{.}}
\]
\end{itemize}

\end{IEEEproof}


The formulae for the packing and the covering radii of a code enable us to characterize the $P$-perfect codes.
\begin{corolario}Let $P=([n],\preceq)$ be a hierarchical poset with $l$ levels, $\mathcal{C}\subseteq \mathbb{F}_q^n$  a linear code and  $t_1 = \min\{i\in[l]: \mathcal{C}_i \neq \{0\}\}$. The code $\mathcal{C}$ is $P$-perfect if, and only if, 
\[\displaystyle \mathcal{C} =\mathcal{C}_{t_1} \oplus \left(\bigoplus_{i=1}^{l-t_1} \mathbb{F}_q^{t_1+i}\right)\mbox{\upshape{}}\]
and $\mathcal{C}_{t_1}$ is a perfect code in $\mathbb{F}_q^{t_1}$ according to the Hamming metric.
\end{corolario}
\begin{IEEEproof}
By definition, a linear code  $\mathcal{C}$ is  $P$-perfect if, and only if, the packing and covering radii are equal.  From Proposition \ref{invariantesmetricos} it follows that $\mathcal{C}$ is $P$-perfect  if, and only if,
\[\mathrm{C}_{ov}(\mathcal{C})-\mathcal{P}(\mathcal{C})=\sum_{i=t_1}^{h-1} n_i + \mathrm{C}_{ov}^H(\mathcal{C}_{{h}}) - \left\lfloor \frac{d_H(\mathcal{C}_{t_1})-1}{2} \right\rfloor = 0\mbox{\upshape{.}}\]
From the definition of $t_1$ and $h$, we have that  $t_1 \leq h$. Since $n_{t_1} >  \left\lfloor \frac{d_H(\mathcal{C}_{t_1})-1}{2}\right\rfloor$ it follows that $h=t_1$ and this implies that $\mathrm{C}_{ov}^H(\mathcal{C}_{{t_1}}) = \left\lfloor \frac{d_H(\mathcal{C}_{t_1})-1}{2}\right\rfloor $. Therefore, $\mathcal{C}_{t_1}$ is a perfect code when considering the Hamming metric on $\mathbb{F}_q^{t_1}$.

Reciprocally, if $\mathcal{C}$ is decomposed as $\mathcal{C} =\mathcal{C}_{t_1} \oplus \left(\bigoplus_{i=1}^{l-t_1} \mathbb{F}_q^{t_1+i}\right)$ and $\mathcal{C}_{t_1}$ is perfect as a code in the Hamming space $\mathbb{F}_q^{t_1}$, then, by  comparing the expressions for the covering and packing radii (items (ii) and (iv) of Proposition \ref{invariantesmetricos}) we find that $\mathcal{P}(\mathcal{C})=\mathrm{C}_{ov}(\mathcal{C})$.
\end{IEEEproof}


In binary spaces, the complement set of $\mathbb{B}_H(u,r)$ is the ball $\mathbb{B}_H(u^c,n-r-1)$ where $u^c_i=0$ if $u_i=1$ and $u^c_i=1$ if $u^c=0$ ($u^c$ is known as the complement vector of $c$). For the Hamming case, as was proved in \cite{frances1997covering}, the covering and the Chebyshev radii are related as follows:

\begin{align*}\label{maz}
	\mathrm{C}_{ov}^H(\mathcal{C}) & =\min\left\{i\in\mathbb{Z} \ : \ \mathbb{F}_q^n= \bigcup_{u\in \mathcal{S}}\mathbb{B}(u,i) \right\} \\
	 & =\max\left\{i\in\mathbb{Z} \ : \ \mathcal{C}\cap \mathbb{B}(u,i)=\emptyset \ \text{ for some } \ u\in\mathbb{F}_q^n \right\}+1\\
	 & = \max\left\{i\in\mathbb{Z} \ : \ \mathcal{C}\subseteq \mathbb{B}(u,n-i-1) \ \text{ for some } \ u\in\mathbb{F}_q^n \right\}+1\\
	 & = \max\left\{n-r-1\in\mathbb{Z} \ : \ \mathcal{C}\subseteq \mathbb{B}(u,r) \ \text{ for some } \ u\in\mathbb{F}_q^n \right\}+1\\
	 & = n-1-\min\left\{r\in\mathbb{Z} \ : \ \mathcal{C}\subseteq \mathbb{B}(u,r) \ \text{ for some } \ u\in\mathbb{F}_q^n \right\}+1\\
	 & = n - \mathcal{R}^H(\mathcal{C}), 
\end{align*}
where $r=n-i-1$. We remark that this identity was previously proved in \cite{karpovsky1981weight}. Out of this, we have the following:

\begin{corolario} \label{q=2}Let $P = ([n],\preceq)$ be a hierarchical poset with $l$ levels. Let $\mathcal{C}\subseteq \F_2^n$ be a binary linear code and $\mathcal{C}_1 \oplus \mathcal{C}_2 \oplus \cdots \oplus \mathcal{C}_l$ be a $P$-canonical decomposition of $\mathcal{C}$ and let $r=\max \{ i\in [l]: \mathcal{C}_i \neq \{ 0  \}\}$. Then, the Chebyshev radius of $\mathcal{C}$ is given by
{
\[\mathcal{R}(\mathcal{C}) = \sum_{i=1}^r n_i - \mathrm{C}_{ov}^H(\mathcal{C}_r).\]}
\end{corolario}
\begin{IEEEproof}
From Proposition \ref{invariantesmetricos}, item (iii),  $\mathcal{R}(\mathcal{C})=\sum_{i=1}^{r-1} n_i + \mathcal{R}^H(\mathcal{C}_{r})$. Since we are considering binary codes, we know that $\mathrm{C}_{ov}^H(\mathcal{C}) = n - \mathcal{R}^H(\mathcal{C})$ and it follows that
\begin{align*}
\mathcal{R}(\mathcal{C}) = n_1+ \cdots +n_{r-1} + n_r - \mathrm{C}_{ov}^H(\mathcal{C}_r).
\end{align*}
\end{IEEEproof}

\begin{remark}
 	Considering the code $\mathcal{C}=\{00,11,22\}\subseteq \F_3^2$, we have that $\mathcal{R}^H(\mathcal{C})=2$ and $C^H_{ov}(\mathcal{C})=1$. It follows that  Corollary \ref{q=2} does not hold for $q\neq 2$. Despite that, from the proof of Corollary \ref{q=2}, it is clear that any relation between the Chebyshev radius and the covering radius in a $q$-ary Hamming space will ensure a similar relation for any hierarchical $P$-space. 
 \end{remark} \vspace{0.2cm}

\subsection{Characterizations of hierarchical poset metrics}

In this section we present some metric properties of codes, each of which is a characterization of hierarchical posets in the sense that, any linear code satisfies the given metric property if, and only if, the metric poset is determined by a hierarchical poset. 
	Despite the fact that most of these properties can be restated by considering any metric in place of the metric Hamming, they are formulated here as related to the poset, not to the metric. Most of those properties are well known in coding theory, sometimes just taken for granted in the case of the Hamming metric. As we shall see in Theorem \ref{main1}, none is obvious by itself, but the metric taken into consideration may turn it so.

We start with one of the main invariants of coding theory, the weight enumerator, which is simply generalized as the $P$-\emph{weight enumerator} of a linear code $\mathcal{C}\subseteq \mathbb{F}_q^n$:
\[W^P_{\mathcal{C}}(X)=\sum_{i=0}^n A_i^P(\mathcal{C})X^i \mbox{\upshape{,}}\] where
\[A_i^P(\mathcal{C})=|\{ c\in \mathcal{C}:\mathrm{wt}_P(c)=i\}|. \] 

In 1961, a remarkable result was presented by Jessie MacWilliams \cite{macwilliams1963theorem},  the MacWilliams Identity,  which relates the weight enumerators of a code $\mathcal{C}$ and its dual $\mathcal{C}^{\perp}$.  
Over the years, relations between weight enumerators of pairs of dual codes  have been explored on a large number of metrics, see \cite{gadouleau2008macwilliams} and \cite{dougherty2002macwilliams}, for example. In the case of poset metrics, to attain such a relation one needs to consider the dual poset: given a poset $P=([n],\preceq_P)$, the \textit{dual poset} is the poset $\overline{P}=([n],\preceq_{\overline{P}})$  defined by the opposite relations, i.e., given $i,j\in [n]$, then 
\[i \preceq_{P}j\iff j\preceq_{\overline{P}}i.\]


\begin{definicao}(\emph{MacWilliams' Identity property})
	A poset $P=([n],\preceq)$ a  Mac\-Williams' Identity if, for any linear code $\mathcal{C}\subseteq \mathbb{F}_q^n$, the $P$-weight enumerator $W^P_{\mathcal{C}}(X)$ of $\mathcal{C}$ determines  the $\overline{P}$-weight enumerator $W^{\overline{P}}_{\mathcal{C}^{\perp}}(X)$ of the dual code $\mathcal{C}^{\perp}$.
\end{definicao}

For poset metrics, the MacWilliams Identity was first investigated in \cite{gutierrez1998macwilliams}. In \cite{classification:macwilliamsidentity} it was proved that the hierarchical posets are the only posets admitting a MacWilliams' identity, furthermore, explicit formulae for those identities was obtained.

Another classical result due to F. J. MacWilliams in the context of  Hamming metric is the MacWilliams Extension theorem, which basically states that the notion of equivalence between codes  may be determined locally. The subject   has been studied in many different contexts, and several papers were devoted to this subject, including \cite{extensiontheoremforfiniterings3}, \cite{dyshko2015extendibility} and \cite{wood1997extension}. 

\begin{definicao}(\emph{MacWilliams' Extension property}\index{MacWilliams Extension})
	A poset $P=([n],\preceq_P)$\emph{ satisfies the MacWilliams Extension property} if for any pair of linear codes $\mathcal{C}$ and $\mathcal{C}'$ and any linear map $t: \mathcal{C}\rightarrow\mathcal{C}'$ preserving the $P$-distance, there is a linear isometry $T\in GL_P(\mathbb{F}_q)$ such that $T|_{\mathcal{C}}=t$.
\end{definicao}

In \cite{extensiontheoremforfiniterings3} it was shown that  hierarchical posets are the unique posets satisfying this   property.
\medskip

 Association schemes, which is a classical structure studied in algebraic combinatorics, came into the picture in \cite{classification:schemes} in order to provide a proof of the MacWilliams identity which is more direct than the one  given in \cite{classification:macwilliamsidentity}.

\begin{definicao}\label{definicaoassociationscheme}Let $X$ be a finite set. Given an integer $m$, consider a set $\mathcal{R} = \{R_0, R_1, \ldots, R_m\}$ of $m+1$ binary relations $R_i$ on $X$ such that $\mathcal{R}$ is a partition of $X\times X$. The pair $(X, \mathcal{R})$ is said to be an \index{association scheme}\emph{association scheme} if the following conditions are satisfied:
	\begin{itemize}
		\item[(i)] $R_0$ is the diagonal, i.e., $R_0=\{(u,u)\in X\times X: u\in X\}$;
		\item[(ii)] $R_i$ is symmetric, i.e., $(u,v)\in R_i$ if, and only if, $(v,u)\in R_i$;
		\item[(iii)] If $(u,v)\in R_k$, then the number of elements $w\in X$ such  that $(u,w)\in R_i$ and $(v,w)\in R_j$ is a constant depending only on $i,j$ and $k$.
	\end{itemize}
\end{definicao}

Whenever we have a metric structure on $X$, a partition of $X\times X$ may be obtained by considering the part $R_i$  to be determined by pairs of points at  distance $i$. As was shown by Delsarte in \cite{delsarte1973algebraic},  this was the key to translate the MacWilliams Identity problem to a distribution problem in the association scheme.

\begin{definicao}We say that a poset $P=([n], \preceq)$\emph{ determines an association scheme} if the pair $(\mathbb{F}_q^n, \mathcal{R}_{d_P})$ is an \index{association scheme} association scheme, where  $\mathcal{R}_{d_P} = \{R_{0,d_P}, R_{1,d_P}, \ldots, R_{n,d_P}\}$ and  
	\[R_{i,d_P} = \{(u,v)\in \mathbb{F}_q^n\times\mathbb{F}_q^n: d_P(u,v)=i\}.\]
\end{definicao}

Conditions to ensure that a poset metric determines  an association scheme  were described in \cite{classification:schemes} and, doing it without considering the $P$-canonical decomposition was a quite difficult task. Other results relating association schemes and poset codes include, for example, conditions over the poset to ensure that the association scheme is self-dual \cite{Barg20141}. 

\medskip

When considering posets consisting of multiple disjoint chains, a MacWilliams' identity is not available. To overcome this difficulty,  the shape of a vector was introduced in \cite{park2010linear} as a refinement of the weight. We give here a very generic definition of a shape:


\begin{definicao}Let $P=([n],\preceq)$ be a poset and $m\in \mathbb{Z}$ be a positive integer. A map $\zeta : \mathbb{F}_q^n\rightarrow \mathbb{Z}^m$ is a \emph{shape mapping} if for every $u,v\in\mathbb{F}_q^n$,
	\[
	\zeta(u)=\zeta(v) \iff \text{ there is } T\in GL_P(\mathbb{F}_q) \text{ such that } T(u)=v.
	\]
\end{definicao}

A shape mapping is ``good" if it has a simple description with $m$ as small as possible. In some cases, we may have $m=1$ and the shape mapping being very simple: $\zeta (u)= \mathrm{wt}_P(u)$. As we shall see, it happens if, and only if, $P$ is hierarchical.

Considering  partial orders as a particular case of directed graphs, they may be studied considering properties of the adjacency matrix, as it is usually done in spectral graph theory, and this is  last structure we are  interested in.

\begin{definicao}Let $P=([n],\preceq)$ be a poset. The \emph{adjacency matrix} $\mathbf{A}_P$  of $P$ is an $n\times n$ matrix with entries satisfying $A_{ij} = 1$ if $i\preceq j$ with $i\neq j$ and zero otherwise.
\end{definicao}
\bigskip
 Now we are in place to present our main result, a vast collection of coding properties, each of  which characterizes the hierarchical posets. 
\bigskip
\begin{teorema}\label{main1}
Let $P = ([n],\preceq)$ be a poset with $l$ levels. Then, $P$ is hierarchical if, and only if, any of the (equivalent) properties below holds: 
\begin{description}
\item[$\mathfrak{P}_0$] Every linear code admits a $P$-canonical decomposition;
\item[$\mathfrak{P}_1$] $P$ admits a MacWilliams' Identity;
\item[$\mathfrak{P}_{2}$] {$P$ satisfies the MacWilliams Extension property;}
\item[$\mathfrak{P}_{3}$] $P$ determines an association scheme;
\item[$\mathfrak{P}_{4}$] The group of linear isometries acts transitively on spheres of a fixed radius, i.e.,
$
\mathrm{wt}(u)=\mathrm{wt}(v) \text{ if, and only if, there is } T\in GL_P(\mathbb{F}_q) \text{ such that } T(u)=v;
$
\item[$\mathfrak{P}_{5}$] The packing radius $\mathcal{P}(\mathcal{C})$ of a linear code $\mathcal{C}$ is a function of its minimum distance $d(\mathcal{C})$;
\item[$\mathfrak{P}_{6}$] The $P$-weight is a shape mapping;
\item[$\mathfrak{P}_{7}$] For each $v\in\mathbb{F}_q^n$, the set $\mathcal{M}(v)$ is contained in $\Gamma_i$ for some $i\in[l]$;
\item[$\mathfrak{P}_{8}$] The entries of the adjacency matrix $\mathbf{A}_P$ satisfy the triangle inequality, i.e., $A_{ij} \leq A_{ik} + A_{kj}$, for all $i,j,k\in [n]$;
\item[$\mathfrak{P}_{9}$] Given two ideals $I,J\subseteq [n]$, then  $|I| = |J|$ if, and only if, they are isomorphic.
\end{description}
\end{teorema}

\bigskip

The characterization given by property $\mathfrak{P}_0$ was actually done in Theorem \ref{thm:canonical_decomposition}; the subscript index $0$ is introduced to stress its key role in the proof of the other statements. Before we proceed to the proof, we recall that some of those statements are known: items $\mathfrak{P}_1$, $\mathfrak{P}_2$, $\mathfrak{P}_3$ and $\mathfrak{P}_4$  are proved in  \cite{classification:macwilliamsidentity}, \cite{extensiontheoremforfiniterings3}, \cite{classification:schemes} and \cite{classification:actstransitively}; item $\mathfrak{P}_5$ is partially proved in \cite{felix_decomposicao_canonica}. Nevertheless, we do present a complete proof for each of these statements, since the $P$-canonical decomposition allows us to do it in an elegant and short way.

We remark that some of these properties are stronger than they might appear at first sight. Consider for example the relation between the packing radius and the minimum distance of a code (Property $\mathfrak{P}_{5}$). 
Even when the minimum distance is known, determining the packing radius in the smallest non-trivial case ($1$-dimensional codes) may be a surprisingly difficult problem: it was proved in \cite{rafno} that, in general, it is an NP-hard problem. We now proceed to the proof.

\bigskip
\begin{IEEEproof}[Proof of  Theorem \ref{main1}] As we have already noted, Theorem \ref{thm:canonical_decomposition} says that  $\mathfrak{P}_0$ is a characterization of hierarchical posets. For the other nine items, we split the proof into two parts: we first prove the ``if" part (arguing by contraposition) for each of the nine statements and, after that, we prove the ``only if" part. 

\medskip

For the \textbf{``if"} part, let us suppose that $P$ is non-hierarchical and denote by $\alpha$ the first level where $P$ ``fails" to be hierarchical, i.e., there is $\alpha \in [l]$ and elements $a \in \Gamma_{\alpha-1}^P$ and  $b \in \Gamma_{\alpha}^P$ such that $a \not\preceq b$. Let us assume that   $\alpha $ is minimal with this condition. In this situation, there must be an element $c \in \Gamma_{\alpha-1}^P$ such that $c \preceq b$ and the structure induced on $\{a,b,c\}$ by $P$ is isomorphic to the tiny counterexample $P_0$. To illustrate this situation consider Figure \ref{pp}. We stress that, on Figure \ref{pp}, the shaded subposet is the one isomorphic to $P_0$.
\begin{figure}[h]
\centering
	\includegraphics[scale=1.3]{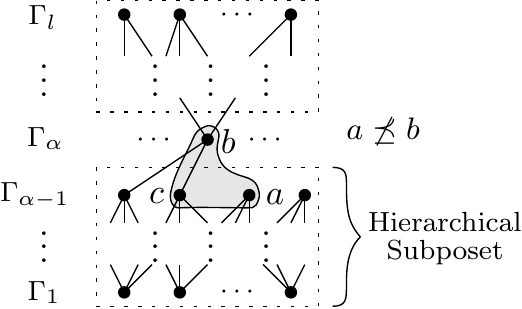}
	\caption{The shaded subposet $\{a,b,c\}$ is isomorphic to $P_0$, the smallest non-hierarchical poset. }\label{pp}
\end{figure}

Considering the subposet $\{a,b,c\}$ we construct two codes that will be a counterexample to many of the listed properties: $\mathcal{C}_1 =span\{ e_b\}$ and  $\mathcal{C}_2 =span \{u\}$, where {$u = \sum_{i\in\langle\{a,b\}\rangle \cap \Gamma_{\alpha-1}^P}e_i$}. The ideals $I_1= \langle \mathrm{supp} (e_b) \rangle=\langle b \rangle $ and $I_2 =\langle \mathrm{supp} (u) \rangle=\langle \{a,b\} \rangle \setminus\{b\} $ generated by the support of each of this (the shaded subposets on Figure \ref{cod}) are hierarchical subposets of $P$.
\begin{figure}[h]
\centering
	\includegraphics[scale=1.3]{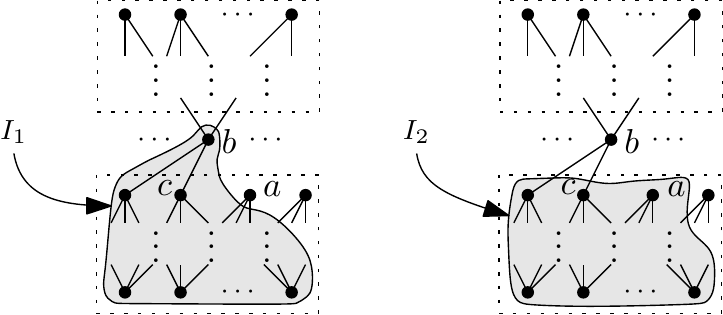}
	\caption{The ideals $I_1= \langle \mathrm{supp}(\mathcal{C}_1) \rangle $ and $I_2= \langle \mathrm{supp}(\mathcal{C}_2) \rangle$.}\label{cod}
\end{figure}

From here on, we consider the level $\alpha$, the vector $u$, the codes $\mathcal{C}_1$ and $\mathcal{C}_2$ and the (hierarchical) ideals  $I_1$ and $I_2$ as defined in the precedent paragraphs. We shall now proceed to prove that properties $\mathfrak{P}_1$ to $\mathfrak{P}_9$ does not hold if the poset $P$ is not hierarchical.

$\mathfrak{P}_1: \;$ Clearly, the $P$-weight enumerators of $\mathcal{C}_1$ and  $\mathcal{C}_2$ are both equal to $1+(q-1)X^{|\langle b\rangle|}$. In order to prove that $W^{\overline{P}}_{\mathcal{C}_1^{\perp}}(X)\neq W^{\overline{P}}_{\mathcal{C}_2^{\perp}}(X)$, it is enough to show that the coefficients $A_{\lambda}^{\overline{P}}(\mathcal{C}_1^{\perp})$ and $A_{\lambda}^{\overline{P}}(\mathcal{C}_2^{\perp})$ are different for some $\lambda\in [n]$. {We recall that $n_i$ is the cardinality of the $i$-th level of $P$, i.e., $n_i = |\Gamma_i^P|$.}

Let $\lambda = n - \sum_{i=1}^{\alpha-2}n_i = \sum_{i=\alpha-1}^{l}n_i$. Since we assume that $P$ is hierarchical up to the level $\alpha -1$, given $v\in\mathbb{F}_q^n$ we have that  $\mathrm{wt}_{\overline{P}}(v)=\lambda$ if, and only if, $\mathcal{M}_{\overline{P}}(v)=\Gamma_{\alpha-1}^P$. Hence,
\begin{equation}\label{eqq001}
	A_{\lambda}^{\overline{P}}(\mathcal{C}_i^{\perp})=|\{c\in\mathcal{C}_i^\perp \ : \ \mathcal{M}_{\overline{P}}(v)=\Gamma_{\alpha-1}^P\}|	
\end{equation}
for $i\in\{1,2\}$. Furthermore, the condition $\mathcal{M}_{\overline{P}}(v)=\Gamma_{\alpha-1}^P$ establishes that $v_i\neq 0$ for every ${i\in\Gamma_{\alpha - 1}^P}$ and that $v_i = 0$ for every $i\in \langle \Gamma_{\alpha-2}^P\rangle$.


Since $\mathcal{C}_1^{\perp} = \{(v_1,\ldots ,v_n)\in\mathbb{F}_q^n: v_b = 0\}$, for every  $v\in\mathcal{C}_1^\perp$ we have that $v_i\in\mathbb{F}_q$ may assume any value for each  $i\in\Gamma_{\alpha}^P\cup\Gamma_{\alpha+1}^P\cup\ldots\cup\Gamma_l^P$. Furthermore, as   $|\Gamma_{\alpha}^P\cup\Gamma_{\alpha+1}^P\cup\ldots\cup\Gamma_l^P|=\lambda-n_{\alpha-1}$, from Equation (\ref{eqq001}), it follows that
\begin{align}\label{1}
A_{\lambda}^{\overline{P}}(\mathcal{C}_1^{\perp}) &= |\{v\in\mathbb{F}_q^n \ : \ \mathcal{M}_{\overline{P}}(v)=\Gamma_{\alpha-1}^P \text{ and } v_b=0\}| \nonumber \\
& = (q-1)^{n_{\alpha-1} }q^{\lambda- n_{\alpha-1} -1} \\ &= (q-1)^tq^{\lambda-n_{\alpha-1} }\frac{(q-1)^{n_{\alpha-1} -t}}{q}. \nonumber
\end{align} 
where $t = n_{\alpha-1}  - \left| \{i \in \Gamma_{\alpha-1}^P : i\preceq b\} \right|$. 

On the other hand, considering the vector $u$, by definition we have that $u_i = 1$ if $i\in\langle\{a, b\} \rangle \cap \Gamma_{\alpha - 1}^P$ and $u_i=0$ if $i\not\in\langle\{a, b\} \rangle \cap \Gamma_{\alpha - 1}^P$. Thus, as $\mathcal{C}_2$ is spanned by $u$, a vector  $v\in\mathbb{F}_q^n$ belongs to  $ \mathcal{C}^{\perp}_2$ if, and only if, $\sum_{i=1}^n v_iu_i =  \sum_{i\in\langle\{a, b\} \rangle \cap \Gamma_{\alpha - 1}^P }v_i = 0$. Hence, from Equation (\ref{eqq001}), 
\[
	A_{\lambda}^{\overline{P}}(\mathcal{C}_2^{\perp}) =|\{v\in\mathbb{F}_q^n \ : \ \mathcal{M}_{\overline{P}}(v)=\Gamma_{\alpha-1}^P \text{ and } \textstyle\sum_{i\in\langle\{a, b\} \rangle \cap \Gamma_{\alpha - 1}^P }v_i =0\}|.
\]
Let $m=|\langle \{a, b\}\rangle \cap \Gamma_{\alpha - 1}^P|$. Given a vector $v$ counted in the determination of $A_{\lambda}^{\overline{P}}(\mathcal{C}_2^{\perp})$, all the $n_{\alpha-1}-m$ coordinates corresponding to  $\Gamma_{\alpha-1}^P \setminus \langle\{a, b\} \rangle$ must be non zero and those non-zero coordinates does not appear as part of the sum $\sum_{i\in\langle\{a, b\} \rangle \cap \Gamma_{\alpha - 1}^P }v_i =0$. Since $n_{\alpha-1}-m=t-1$ and, as before, the $\lambda-n_{\alpha-1}$ coordinates $\Gamma_{\alpha}^P\cup\Gamma_{\alpha+1}^P\cup\ldots\cup\Gamma_l^P$ may assume any value, 
\[
	A_{\lambda}^{\overline{P}}(\mathcal{C}_2^{\perp}) =(q-1)^{t-1}q^{\lambda-n_{\alpha-1}}|\{v\in\mathbb{F}_q^m \ : \textstyle\sum_{i=1}^m v_i =0 \text{ and } v_i\neq 0\}|.
\]
The number of solutions of $\sum_{i=1}^m v_i =0$,  with $v_i\in\mathbb{F}_q\setminus\{0\}$, denoted by $S_m$, is the number of $m$-\emph{compositions} over $\mathbb{F}_q$ (in the combinatorial sense, see \cite{stanley}). It is known by \cite{numerodecomposicoesdezero} that 
 \[S_{m} =\frac{(q-1)^{m}+(-1)^{m}(q-1)}{q}.\]
Therefore,
\begin{equation}\label{2}
A_{\lambda}^{\overline{P}}(\mathcal{C}_2^{\perp}) =(q-1)^{t-1} q^{\lambda-{n_{\alpha-1} }}S_{ m }. 
\end{equation}
Since $S_{m} \neq \frac{(q-1)^{m}}{q}$, considering equations (\ref{1}) and (\ref{2}) we find that  $A_{\lambda}^{\overline{P}}(\mathcal{C}_1^{\perp})\neq A_{\lambda}^{\overline{P}}(\mathcal{C}_2^{\perp})$ hence the $\overline{P}$-weight enumerators  of $\mathcal{C}_1^{\perp}$ and  $\mathcal{C}_2^{\perp}$ are different.



$\mathfrak{P}_2: \;$   For every non-zero $\lambda\in\mathbb{F}_q$, the map $t: \mathcal{C}_1 \rightarrow \mathcal{C}_2$ defined by $t(\lambda \!\cdot \!e_b) = \lambda \!\cdot\! u$ is a linear $P$-isometry between $\mathcal{C}_1$ and $\mathcal{C}_2$. This map cannot be extended to a linear $P$-isometry of $\mathbb{F}_q^n$, indeed, item (\textit{i}) of Theorem \ref{idealprimo} ensures that if $T\in GL_P(\mathbb{F}_q)$, then $\langle \mathrm{supp}(T(e_b))\rangle$ is a prime ideal, but $T(e_b)= t(e_b)=u$ and $\langle \mathrm{supp}( u)\rangle$ is clearly not             prime.

$\mathfrak{P}_3: \;$  Conditions (i) and (ii) in  the definition of an association scheme are satisfied independently of the metric. Thus, assuming that $P$ is not hierarchical, we provide a general counterexample to prove that the condition (iii) in Definition \ref{definicaoassociationscheme} is not satisfied. Let $s = \sum_{i=1}^{\alpha-2} n_i$. By definition $(0,e_b)\in R_{\mathrm{wt}(e_b),d_P}$.  Moreover, the set
\[\{z\in\mathbb{F}_q^n : (0,z)\in R_{s+1, d_P} \mbox{\upshape{ and }} (e_b,z)\in R_{\mathrm{wt}(e_b)-1, d_P}\}\]
is empty, because $d_P(0,z) = s+1$ implies $z_b=0$ and $d_P(e_b,z)=\mathrm{wt}(e_b)-1$ implies $z_b=1$. On the other hand, by the choice of $u$, we have that $(0,u)\in R_{\mathrm{wt}(e_b),d_P}$. However, the set  \[\{z\in\mathbb{F}_q^n : (0,z)\in R_{s+1, d_P} \mbox{\upshape{ and }} (u,z)\in R_{\mathrm{wt}(e_b)-1, d_P}\}\] is not empty, since $d_P(0,e_a) =s+1$ and $d_P(u,e_a)= \mathrm{wt}(e_b) - 1$. It follows that $P$ does not determine an association scheme.

%

$\mathfrak{P}_4: \;$   Since $\mathrm{wt}(e_b)=\mathrm{wt}(u)$ and  $\mathrm{supp}(e_b)$ generates a prime ideal while $\mathrm{supp}(u)$ does not, it follows, from item (i) in Theorem \ref{idealprimo}, that $e_b$ cannot be mapped into $u$ by any $P$-isometry $T\in GL_P(\mathbb{F}_q)$. 


$\mathfrak{P}_5: \;$ Proposition \ref{just} ensures that in order to obtain the packing radius of both $\mathcal{C}_1$ and $\mathcal{C}_2$, we may puncture $\mathcal{C}_1$ and $\mathcal{C}_2$ on the coordinates corresponding to the ideals $I = \langle b \rangle_P$ and $J = \langle a, b \rangle_P\setminus \{b\}$, respectively. Furthermore, since both $I$ and $J$ are hierarchical posets (shaded posets in Figure \ref{cod}), the packing radius of $\mathcal{C}_1$ and $\mathcal{C}_2$ are determined in item (\textit{ii}) of Proposition \ref{invariantesmetricos}: 
\[\mathcal{P}(\mathcal{C}_2)  =\sum_{i=1}^{\alpha-2} n_i + \left\lfloor\frac{|\Gamma_{\alpha -1}^P\cap \langle b \rangle |}{2}\right\rfloor\]
and
\[\mathcal{P}(\mathcal{C}_1) =\sum_{i=1}^{\alpha-2} n_i + |\Gamma_{\alpha -1}^P\cap \langle b \rangle_P|.\]
Since $|\Gamma_{\alpha -1}^P\cap \langle b \rangle_P |>0$ we find that $\mathcal{P}(\mathcal{C}_1)> \mathcal{P}(\mathcal{C}_2)$ but, by construction, $d(\mathcal{C}_1)=d(\mathcal{C}_2)$. 

$\mathfrak{P}_6: \;$ 
If the $P$-weight is a shape mapping, then the group of linear isometries acts transitively on spheres of a fixed radius, but, as seen in the proof of $\mathfrak{P}_4$, the vectors $u$ and $e_b$ have the same weight and are not in the same orbit, since there is no linear $P$-isometry $T$ such that $T(u)=e_b$.

$\mathfrak{P}_7: \;$ If we take $v = e_a + e_b$, then we have that $\mathcal{M}(v)\cap \Gamma_{\alpha-1}^P= \{a\}$ and $\mathcal{M}(v)\cap \Gamma_{\alpha}^P=\{b\}$.

$\mathfrak{P}_8: \;$ 
Since $b \in \Gamma_{\alpha}^P$ and $c\in \Gamma_{\alpha-1}^P$ satisfy $c \preceq b$, it follows that $A_{cb}= 1$. On the other hand, $A_{ca} = A_{ab}= 0$. Therefore, the triangle inequality does not hold.

$\mathfrak{P}_9: \;$ The ideals $I = \langle b \rangle$ and $J = \{a\} \cup \left(I\backslash \{b\}\right)$ have same cardinality. However, there is no isomorphism between $I$ and $J$, because $I$ is prime while $J$ is not.

\bigskip

To prove the \textbf{``only if"} part, consider $P$ to be a hierarchical poset with $l$ levels and let $n_i=|\Gamma_i^P|$ for every $i\in[l]$. Let $\mathcal{C}\subseteq \mathbb{F}_q^n$ be a linear code. Without loss of generality (ensured by Theorem \ref{thm:canonical_decomposition}), we assume that  $\mathcal{C}=\mathcal{C}_1\oplus\cdots\oplus \mathcal{C}_l$, with $\mathrm{supp}(\mathcal{C}_i)\subseteq \Gamma_i^P$ and $i\in[l]$.

$\mathfrak{P}_1: \;$ The weight enumerator of $\mathcal{C}$ is given by 
\begin{align*}
W^P_{\mathcal{C}}(X)=1 + W_{\mathcal{C}^*_1}(X)+ X^{s_1}W_{\mathcal{C}_2^*}(X)&|\mathcal{C}_1|+ \cdots \\&+X^{s_{l-1}}W_{\mathcal{C}_l^*}(X)|\mathcal{C}_1||\mathcal{C}_2|\cdots|\mathcal{C}_{l-1}|,
\end{align*}
where $\mathcal{C}_i^* = \mathcal{C}_i\setminus \{0\}$,  $W_{\mathcal{C}_i^*}(X)$ is the Hamming weight enumerator of $\mathcal{C}_i^*\subseteq \F_q^{n_i}$ and $s_j = \sum_{i=1}^jn_i$.

Let us define $D_i=\{v\in \mathcal{C}_i^{\perp}: \mathrm{supp}(v)\subseteq \Gamma_i^P \} $. It is clear that each $D_i$ is a vector subspace contained in $\mathcal{C}^{\perp}$ and, from simple dimensionality reasoning, it follows that $\mathcal{C}^{\perp} = D_1 \oplus D_2 \oplus\cdots \oplus D_l$. Moreover, this is, by itself, a $\overline{P}$-canonical decomposition of $\mathcal{C}^\perp$. Thus, the $\overline{P}$-weight enumerator of $\mathcal{C}^\perp$ may be written as 
\begin{align}\label{eqMac}
W^{\overline{P}}_{\mathcal{C}^{\perp}}(X) = 1 + W_{D_l^{*}}(X)+ X^{\overline{s_1}}W_{D_{l-1}^*}(X)&|D_l|+ \cdots \\
&+X^{\overline{s_{l-1}}}W_{D_1^*}(X)|D_2||D_3|\cdots|D_l|,\nonumber
\end{align}
where $D_i^*=D_i\setminus\{0\}$, $W_{D_i^*}(X)$ is the Hamming weight enumerator of $D_i^*\subseteq \mathbb{F}_q^{n_i}$ and $\overline{s_j}=\sum_{i=l-j+1}^{l}n_i$. For each $i\in[l]$, the punctured codes obtained by puncturing $\mathcal{C}_i$ and $D_i$ on $[n]\setminus\Gamma_i^P$ have the same Hamming weight enumerator of $\mathcal{C}_i$ and $D_i$, respectively. Furthermore, the punctured codes are dual to each other and then, the classical MacWilliams identity ensures that the parcel $W_{D_i^*}(X)$ of the weight enumerator of $\mathcal{C}^\perp$ is uniquely determined by the parcel $W_{\mathcal{C}_{i}^*}(X)$ of the weight enumerator of $\mathcal{C}$. It follows that the $\overline{P}$-weight enumerator of $\mathcal{C}^\perp$ may be fully determined once the  ${P}$-weight enumerator of $\mathcal{C}$ is given.

$\mathfrak{P}_2: \;$ Let $\mathcal{C}^\prime = \mathcal{C}_1^\prime \oplus \cdots \oplus \mathcal{C}_l^\prime$ be a linear code on $\mathbb{F}_q^n$ and $t: \mathcal{C} \rightarrow \mathcal{C}'$ be a linear map that preserves the $P$-weight. Since $t$ is assumed to preserve the $P$-weight, given $c_i\in\mathcal{C}_i$, we have that $\displaystyle t(c_i)=t_i(c_i) + f_i(c_i)$ where $f_i: \mathcal{C}_i \rightarrow \bigoplus_{j<i}\mathcal{C}_j'$, $t_i:\mathcal{C}_i \rightarrow \mathcal{C}_i'$ and  $t_i$ is a $P$-isometry. Moreover, since $t$ is assumed to be a linear map, we have that both $t_i$ and $f_i$ are linear. The linearity of $t$ ensures that, given $c=c_1+\cdots+c_l$ with $c_i\in\mathcal{C}_i$, we have that
\[
t(c)=\sum_{i=1}^l t_i(c_i) + f_i(c_i).
\]
Since $P$ is hierarchical and $\mathrm{supp}(\mathcal{C}_i)$ and $\mathrm{supp}(\mathcal{C}_i')$ are subsets of $\Gamma_i$, it follows that $t_i$ is also a linear isometry according to the Hamming metric. The classical MacWilliams Extension ensures the existence of a linear isometry $T_i: \mathbb{F}_q^{n_i}\rightarrow \mathbb{F}_q^{n_i}$ such that $T_i\vert_{\mathcal{C}_i} = t_i$. Because $P$ is hierarchical and $\mathrm{supp}(\mathbb{F}_q^{n_i})=\Gamma_i^P$, the map $T_i$ is a linear isometry according to the poset metric $P$. For each $i\in [l]$, let us consider $\mathbb{F}_q^{n_i} $ as the direct sum $\mathbb{F}_q^{n_i} = \mathcal{C}_i \oplus W_i$ and define $F_i:\mathbb{F}_q^{n_i}\rightarrow\mathbb{F}_q^{n}$  as the linear map determined by $F_i(u+w)=f_i(u)$ for $u\in \mathcal{C}_i $ and $w\in W_i $. Since each $v\in\mathbb{F}_q^n$ may be uniquely decomposed as $v=v_i+\cdots+v_l$ with $\mathrm{supp}(v_i)\subseteq\Gamma_i^P$ for every $i\in [l]$, the map
\begin{equation*}
	\begin{tabular}{cccc}
	$T : $& $\mathbb{F}_q^n$     & $\rightarrow$  & $\mathbb{F}_q^n$ \\
    & $v_1+ \cdots + v_l$  &$\mapsto$       & $\sum_{i=1}^l T_i(v_i)+F_i(v_i)$
\end{tabular}
\end{equation*}
is well defined. Furthermore, by construction, $T$ is a linear $P$-isometry satisfying $T\vert_{\mathcal{C}}=t$.

$\mathfrak{P}_3:$  Since $P$-distances are invariant by translations, in order to prove the condition (iii) of Definition \ref{definicaoassociationscheme}, when considering a pair of vectors $(u,v)\in R_k$, we may assume $u=0$. So, it is enough to show that the cardinality of the sets $S_{i,j}^v=\{w\!\in\!\mathbb{F}_q^n\!:\! d_P(v,w) = j\mbox{\upshape{ and }}\mathrm{wt}(w)= i \}$ does not depend on the choice of $v\in\mathbb{F}_q^n$ but only on $\mathrm{wt}(v)=k$. Consider $u,v\in\mathbb{F}_q^n$ such that $\mathrm{wt}(u)=\mathrm{wt}(v)$. By the MacWilliams Extension property ($\mathfrak{P}_2$), there is $T\!\in GL_P(\F_q)$ such that $T(u)=v$. Hence, $w \in S_{i,j}^u$ if, and only if, $T(w)\! \in S_{i,j}^{v}$. Therefore, $|S_{i,j}^u|=|S_{i,j}^v|$.

$\mathfrak{P}_4: \;$ Given $r\in\mathbb{Z}$ and $u, v\in\mathbb{S}_P(0, r)$, the map defined by $t(\lambda u ) = \lambda v$ for every $\lambda\in\mathbb{F}_q$ is a linear map between the spaces generated by $u$ and $v$ preserving the $P$-weight.  The MacWilliams Extension property ($\mathfrak{P}_2$) ensures that $t$ may be extended to a map $T\in GL_P({\mathbb{F}_q})$. Since $T(u)=v$, the group $GL_P({\mathbb{F}_q})$ acts transitively on $\mathbb{S}_P(0, r)$.



$\mathfrak{P}_5: \;$ Follows direct from the formula of the packing radius given in Proposition \ref{invariantesmetricos}.

$\mathfrak{P}_6: \;$ Given $v\in\mathbb{F}_q^n$, from Property $\mathfrak{P}_4$ we have that 
\[\{T(v): T\in GL_P(\mathbb{F}_q)\} = \mathbb{S}_P(0, \mathrm{wt}_P(v))\mbox{\upshape{,}}\] and so the $P$-weight is a shape mapping.


$\mathfrak{P}_7: \;$ 
Given $v\in\mathbb{F}_q^n$, the elements of $\mathcal{M}(v)$, being maximals, are not comparable to each other. On the other hand, since $P$ is hierarchical, two elements belonging to different levels are always comparable. Hence, $\mathcal{M}(v)\subseteq \Gamma_s$ for some $s\in[l]$.

$\mathfrak{P}_8: \;$ If $A_{ij}=0$, there is nothing to be proved. Let us assume that $A_{ij}=1$. This implies that $i\preceq j$ and $i\neq j$. We need to prove that either $A_{ik} =1$ or $ A_{kj}=1$. Given $k\in [n]$, if $k=i$ then $A_{kj}=1$. Suppose $k\neq i$, then $k$ and $i$ are either comparable or not comparable. In the first case, if $i\preceq k$, then $A_{ik}=1$, on the other hand, if $k\preceq i$, then $k\preceq j$ since $i\preceq j$, hence $A_{kj}=1$. For the second case, $A_{kj}=1$ because $i\preceq j$ and the elements $i$ and $k$ belong to the same level of $P$. 


$\mathfrak{P}_9: \;$  Since $P$ is hierarchical, given two ideals $I, J \subseteq [n]$, it follows that $|I| = |J|$ if, and only if, $|\mathcal{M}(I)| = |\mathcal{M}(J)|$ and $\mathcal{M}(I), \mathcal{M}(J)\subseteq \Gamma_r$, for some $r \in[l]$. Hence, if $|I|=|J|$, then  
\[I =  \left(\bigcup_{j=1}^{r-1}\Gamma_j\right)\cup \mathcal{M}(I) \text{ and}\;\; J = \left(\bigcup_{j=1}^{r-1}\Gamma_j\right)\cup \mathcal{M}(J)\]
for some $r\in[l]$. Consider a map $\phi: I \rightarrow J$ determined as follows:  $\phi(\mathcal{M}(I)) = \mathcal{M}(J)$ is any bijection and $\phi|_{\bigcup_{j=1}^{r-1}\Gamma_j}$ is the identity function. By the construction of $\phi$ and since $P$ is assumed to be hierarchical, we have that $\phi$ is an isomorphism between the ideals $I$ and $J$.

\end{IEEEproof}

To conclude the work, we make explicit the MacWilliams identity for hierarchical posets, which existence was proved in Theorem \ref{main1}.

In order to simplify the notation we consider the polynomial
\begin{equation}\label{auxiliarypolynomial}
A_{\mathcal{C}}(X)=\left(1 + (q-1)X \right)^n W_{\mathcal{C}}\left(\frac{1-X}{1-(q-1)X}\right)-1,
\end{equation}
where $W_{\mathcal{C}}(X)$ is the weight enumerator of $\mathcal{C}$ relative to the Hamming metric.

\begin{corolario} [The MacWilliams Identity] Let $P$ be a hierarchical  poset with $l$ levels and $n_i = |\Gamma_i|$. Let $\mathcal{C}\subseteq\mathbb{F}_q^n$ be a linear code, $\mathcal{C}_1 \oplus \cdots \oplus \mathcal{C}_l$ its canonical decomposition and $k_i = dim(\mathcal{C}_i)$. Then, the $\overline{P}$-weight enumerator of its dual code is given by
	\begin{align*}
	W^{\overline{P}}_{\mathcal{C}^{\perp}}&(X) = &1 + \frac{1}{q^{k_l}}A_{\mathcal{C}_l}\left(X\right)+ X^{\overline{s_1}}\frac{q^{n_l-k_l}}{q^{k_{l-1}}}A_{\mathcal{C}_{l-1}}\left(X\right)+ \cdots +&X^{\overline{s_l}}\frac{q^{n_2-k_2}q^{n_3-k_3}\cdots q^{n_l-k_l}}{q^{k_1}}A_{\mathcal{C}_1}\left(X\right),
	\end{align*}
	where $\overline{s_j}=\sum_{i=l-j+1}^{l}n_i$.
\end{corolario} 
\begin{IEEEproof}
	From the classical MacWilliams Identity we have that 
	\begin{eqnarray}\label{3}
W_{D_i}(X) = \frac{1}{q^{k_i}}A_{\mathcal{C}_i}(X) + 1,
	\end{eqnarray}
where $D_i=\{v\in \mathcal{C}_i^{\perp}: \mathrm{supp}(v)\subseteq \Gamma_i \}$. The identity follows straightforward from equations (\ref{eqMac}) and (\ref{3}).
	
%
%
\end{IEEEproof}

%
%

    \section*{Acknowledgment}
The first two authors would like to thanks the National Council for Scientific and Technological Development (CNPq, Brazil) for the financial support. All the authors were partially supported by S\~{a}o Paulo Research Foundation (FAPESP), through grants  2013/25977-7, 2015/11286-8 and 2016/01551-9. 

    \bibliographystyle{IEEEtran}
    \bibliography{biliog}

\begin{IEEEbiographynophoto}{Roberto A. Machado}
received a B.Sc. in Mathematics (2013), from University of Lavras.
He holds a master degree from the Institute of Mathematics, Statistics and Scientific Computing of the University of Campinas, where he is currently a PhD candidate. His current research interests include poset codes.
 \end{IEEEbiographynophoto}
\begin{IEEEbiographynophoto}{Jerry A. Pinheiro}
received a B.Sc. in Computer Science (2006) and a B.Sc. in Mathematics (2008), from Higher Education Center of Foz do Igua\c cu and State University of Western Paran\'a respectively. He received a M.Sc. and a Ph.D. degrees in 2011 and 2016 respectively, both from University of Campinas, Brazil. He is currently a postdoc in the Institute of Mathematics, Statistics and Scientific Computing of the University of Campinas. His current research interests include metrics in coding theory and algebraic combinatorics.
\end{IEEEbiographynophoto}
\begin{IEEEbiographynophoto}{Marcelo Firer}
received B.Sc. and M.Sc. degrees in 1989 and 1991 respectively, from University of Campinas, Brazil, and a Ph.D. degree from the Hebrew University of Jerusalem, in 1997, all in Mathematics.
He is currently an Associate Professor of the University of Campinas. His research interests include coding theory, action of groups, semigroups and Tits buildings.
\end{IEEEbiographynophoto}
\vfill
\end{document}